\documentclass[11pt,usenatbib]{mn2e}
\usepackage{graphicx}

\usepackage{hyperref}

\newcommand{\gtrsim}{\ga}

\newcommand{\apj}{ApJ}

\newcommand{\apjs}{ApJS}
\newcommand{\mnras}{MNRAS}

\usepackage{color}

\definecolor{grey}{rgb}{0.7,0.7,0.7}

\begin{document}
\topmargin -0.5in 

\title[High-redshift galaxies and low-mass stars]{High-redshift galaxies and low-mass stars}

\author[Stephen M. Wilkins et al.]  
{
Stephen M. Wilkins$^{1}$\thanks{E-mail: s.wilkins@sussex.ox.ac.uk}, Elizabeth R. Stanway$^2$, Malcolm N. Bremer$^3$\\
$^1$\,Astronomy Centre, Department of Physics and Astronomy, University of Sussex, Brighton, BN1 9QH, U.K. \\
$^2$\,Department of Physics, University of Warwick, Coventry, CV4 7AL, UK \\
$^3$\,H H Wills Physics Laboratory, Tyndall Avenue, Bristol, BS8 1TL, UK
}
\maketitle 

\begin{abstract}

The sensitivity available to near-infrared surveys has recently allowed us to probe the galaxy population at $z\approx 7$ and beyond. The existing
{\em Hubble} Wide Field Camera 3 (WFC3) and the Visible and Infrared Survey Telescope for Astronomy (VISTA) Infrared Camera (VIRCam) instruments allow
deep surveys to be undertaken well beyond one micron - a capability
that will be further extended with the launch and commissioning of the
{\em James Webb Space Telescope (JWST)}. As new regions of parameter
space in both colour and depth are probed new challenges for
distant galaxy surveys are identified. In this paper we present an
analysis of the colours of L and T dwarf stars in widely used
photometric systems. We also consider the implications of the
newly-identified Y dwarf population - stars that are still cooler and
less massive than T dwarfs for both the photometric selection and
spectroscopic follow-up of faint and distant galaxies.  We highlight
the dangers of working in the low-signal-to-noise regime, and the
potential contamination of existing and future samples. We find that {\em Hubble}/WFC3 and VISTA/VIRCam $Y$-drop selections targeting galaxies at $z\sim7.5$ are vulnerable to contamination
from T and Y class stars. Future observations using {\em JWST}, targeting the $z\sim7$ galaxy population, are also likely to prove difficult without deep medium-band observations. We demonstrate that single emission line detections in
typical low signal-to-noise spectroscopic observations may also be
suspect, due to the unusual spectral characteristics of the cool dwarf
star population.
\end{abstract} 


\section{Introduction}

The Lyman break technique is one of the most widely used methods to identify distant galaxies in
extragalactic astronomy, and also one of the most successful. The strong
spectral break at 1216\AA\ in the rest frame of distant galaxies,
caused by hydrogen absorption in the intergalactic medium, provides a
clear and distinctive indication of their redshift
\citep{1990ApJ...357L...9G}. The identification of this feature
through photometric imaging has proven highly efficient when applied
at $z\sim3$ \citep{1999ApJ...519....1S,1996ApJ...462L..17S}. At this
redshift, colours in photometric filters bracketing the Lyman $\alpha$
feature and Lyman limit can be used to select star-forming galaxies
with negligible contamination from lower redshift systems
\citep{2003ApJ...592..728S}. Observations in filters shortwards of the
break will show the Lyman break galaxy (LBG) in question to be faint
or perhaps entirely absent, leading to the widely used term `dropout'
to describe higih-redshift galaxies that drop below the observational
limit in a particular filter and at a particular redshift.

However, as the method is extended to higher redshifts, difficulties
can arise. Balmer break systems and strong emission line galaxies at
intermediate redshifts can contaminate so-called `dropout' samples,
showing similar strong flux decrements in filters that lie either side
of their primary spectral features.

Similarly, cool and low mass dwarf stars can have extreme and unusual
colours. Mid- to late M-class stars are brighter at red wavelengths
than in the blue, giving them colours that imitate the $R-$ or
$V-$dropout colour that is used to identify candidate $z\sim4.5-5.5$
galaxies \citep{2009ApJ...695.1591P,2008MNRAS.384..348S}. Cooler L- and
early T-class stars have proven a contaminant in $I$-band dropout
samples targeted at selecting galaxies at $z\sim6$
\citep{2004ApJ...607..704S,2008A&A...488..181C}. 

Various methods can be used to identify and exclude these
contaminants.  Galaxies at $z\sim4-6$ are usually resolved in deep
{\em Hubble Space Telescope} ({\em HST}) optical data \citep{2004ApJ...600L.107F},
allowing morphological separation for all but the faintest
sources. From the ground, or for very faint galaxies, the task is more
challenging. Where morphological information is not available, it is
possible to identify M- and L-dwarf stars through their red colours in
the near-infrared \citep[see][]{2008MNRAS.384..348S}. For relatively
bright $z\sim5-6$ galaxies it may also be possible to obtain
spectroscopy which reaches the continuum level \citep[e.g.][]{2003ApJ...593..630L}. M- and L-stars show
abrupt breaks in their spectra, similar to those in Lyman break
galaxies, but the continuum longwards of the first detected break will
be interrupted by deep molecular absorption lines and relatively red
where not absorbed, allowing them to be distinguished from true high
redshift objects
\citep[e.g.][]{2004ApJ...607..704S,2010MNRAS.409.1155D}.

As Lyman break-based selection methods shift to identifying $z-$ and
$Y-$drop galaxies at $z\sim7$ and 8 respectively \citep[e.g.][]{2010MNRAS.409..855B,2010MNRAS.403..938W,2011MNRAS.411...23W,2011MNRAS.414.1455L,2013MNRAS.429..150L}, the photometric
bands used to identify target sources lie in the infrared. Current
observations \citep[e.g. the CANDELS survey, ][]{2011ApJS..197...36K}
are probing a region of parameter space never before studied, reaching
unprecedented depths in near-infrared imaging. However, whenever new
parameter space is explored, a problem arises: how can one assess the
reliability of the resulting observations and the potential for new
contaminant populations?

Near-infrared observations, even with {\em HST}, cannot reliably
separate stars from compact galaxies by morphology without leading to incompleteness
in the galaxy samples. Observations significantly longwards of the
break are highly challenging if not impossible, requiring imaging at
$>3$\,$\mu$m at depths beyond the reach of the {\em Spitzer Space
  Telescope} for typical sources. Spectroscopy has also proved
extremely challenging and expensive in observing time, with
spectrographs on 8-10m telescopes unable to reach the continuum
level at a typical magnitude $J_{AB}\sim27.5$. Since deep
extragalactic surveys probe magnitude limits not reached by previous
observations, models of the expected number of T-dwarf stars (known to
meet $z>7$ selection criteria) to a given depth and in a given area
are based on an understanding of Galactic structure, extrapolated well
beyond the observed sources in the solar neighbourhood. To add to the
complication, T-dwarfs themselves have been found to be a diverse and
unpredictable population \citep{2013ApJS..205....6M}.

In addition, the sequence of known cool, low-mass stars -- from M, to
L and through to T -- has recently been extended still further. A new,
cooler class of brown dwarfs, Y-dwarfs, have been identified in near-infrared
data \citep{2012ApJ...753..156K,2011ApJ...743...50C}.  The list of
  known stars in this class is growing, but the faint absolute
  magnitudes of these sources still limits studied Y- and late
  T-dwarfs to stars in the immediate solar neighbourhood
  \citep[$<20$pc, ][]{2013arXiv1308.5372K,2013ApJ...762..119M}.

Given the strong molecular absorption in the spectra of the cool
atmospheres of these brown dwarfs, they are essentially undetectable
in the optical, but potentially bright in infrared bands. This may
well place them in the selection windows used to select $z>7$ galaxies
\citep[e.g.][]{2010ApJ...725.1587B,2011MNRAS.411...23W,2011MNRAS.414.1455L}.
Their precise
colours, though, will depend on the strength and spectral location of
the absorption regions and thus on the composition and temperature of the
stellar atmospheres \citep[e.g. ][]{2012ApJ...756..172M}. 

In this paper we consider in detail the potential of the late T and Y-dwarf
stellar population to contaminate $z>7$ candidate galaxy samples and the
extent to which this can be mitigated in deep samples. In section \ref{sec:browndwarfs} we discuss the known and modelled properties of the brown dwarf population. In section \ref{sec:photo} we consider their photometric colours and potential impact on higih-redshift photometric samples. In section \ref{sec:spectro} we consider the potential of these sources to contaminate spectroscopic samples. Finally, in section \ref{sec:conc} we present our conclusions.
Throughout, photometric magnitudes are given in the AB system.

\section{Brown dwarfs}\label{sec:browndwarfs}

\begin{table}\label{tab:LTsources}
\caption{Original source references for the L and T-dwarf spectra from the SpeX Prism Spectral Library used in this work.}
\centering
\begin{tabular}{ll}
\hline
Type & Reference \\
\hline
L1 & Kirkpatrick et al. (2010) ApJS, 190, 100  \\
L3 & Burgasser (2007) ApJ, 659, 655  \\
L4 & Kirkpatrick et al. (2010) ApJS, 190, 100  \\
L4.5 & Chiu et al. (2006) AJ 131, 2722  \\
L6 & Cruz et al. (2004) ApJ, 604, L61  \\
L8 & Burgasser (2007) ApJ, 659, 655  \\
L9 & Burgasser et al. (2006) ApJ, 637, 1067  \\
\hline
T0 & Looper, Kirkpatrick, and Burgasser (2007) AJ, 134, 1162  \\
T1 & Burgasser et al. (2004) AJ, 127, 2856  \\
T2 & Burgasser et al. (2004) AJ, 127, 2856  \\
T3 & Burgasser et al. (2004) AJ, 127, 2856  \\
T4 & Burgasser et al. (2004) AJ, 127, 2856  \\
T5 & Burgasser et al. (2004) AJ, 127, 2856  \\
T6 & Burgasser et al. (2006) ApJ, 639, 1095  \\
T7 & Burgasser et al. (2006) ApJ, 639, 1095  \\
T8 & Burgasser et al. (2004) AJ, 127, 2856  \\
\hline
\end{tabular}
\label{tab:params}
\end{table}

\begin{figure}
\centering
\includegraphics[width=20pc]{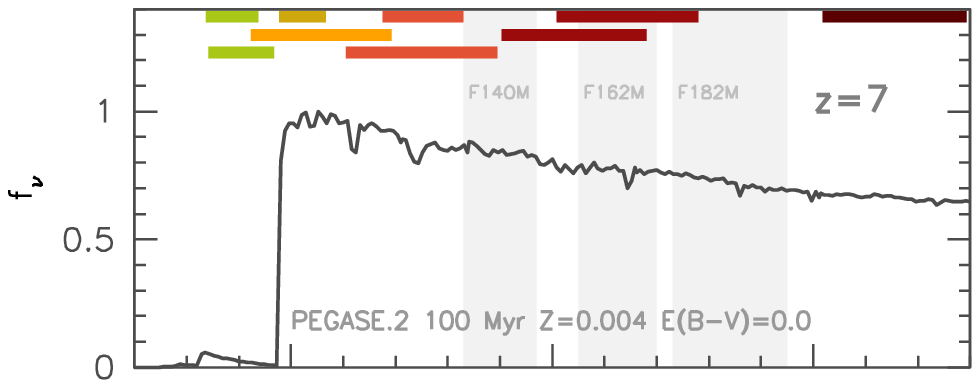}
\vskip -40pt
\includegraphics[width=20pc]{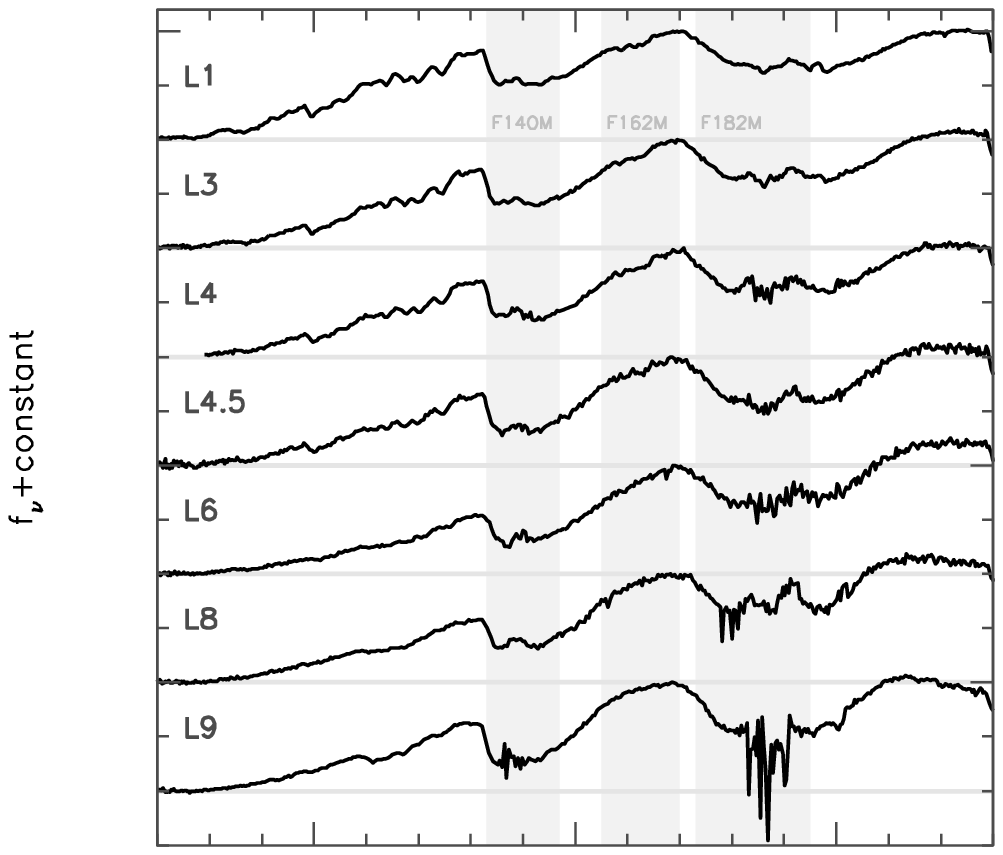}
\vskip -40pt
\includegraphics[width=20pc]{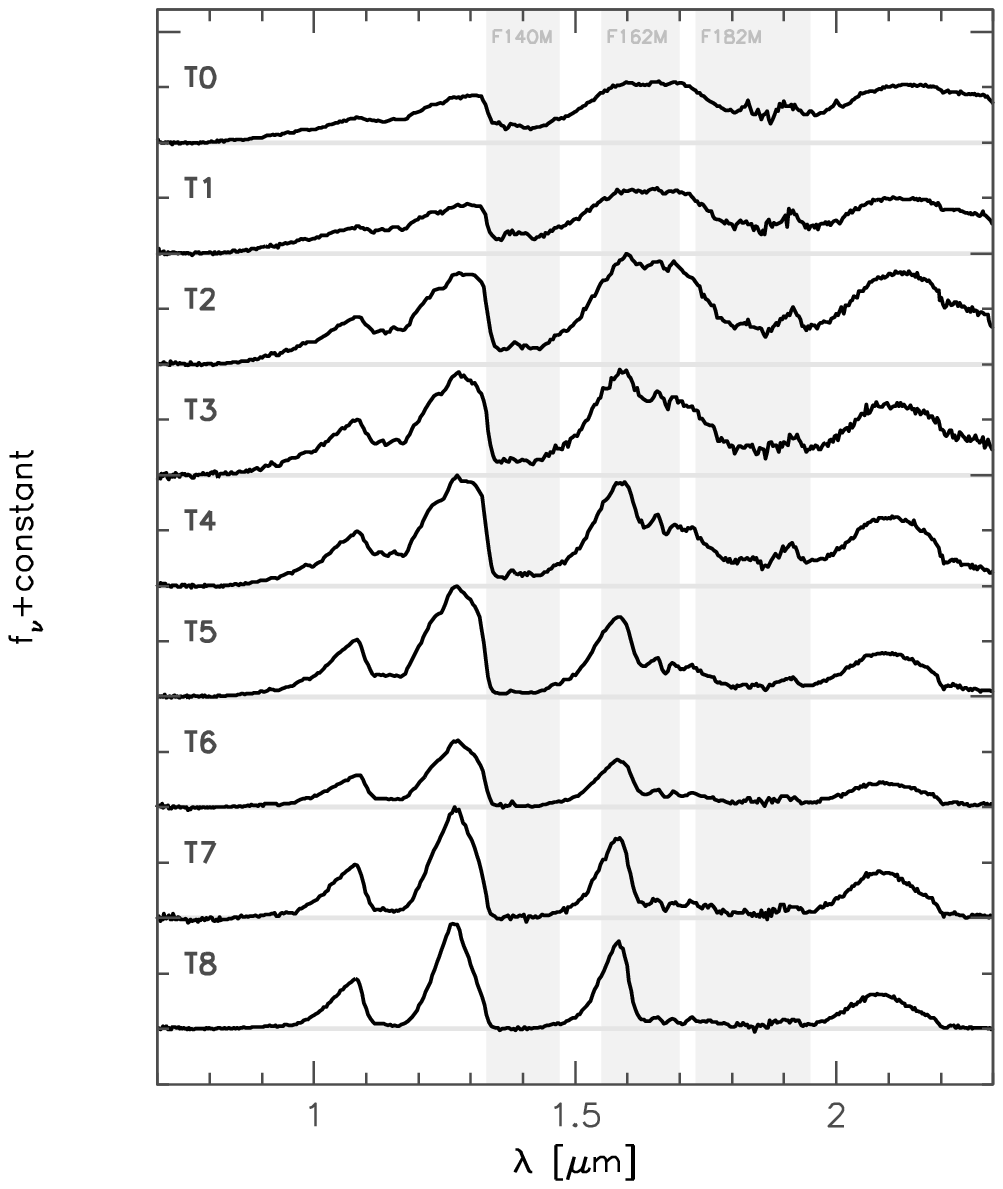}
\caption{{\em top - } Template (using the {\sc PEGASE.2} stellar population synthesis model) dust-free star-forming galaxy at $z=7$. {\em middle and bottom -} Spectra of L and T-dwarf spectral standards from the SpeX Prism Spectral Library. See Table \ref{tab:LTsources} for references. The three grey bands show the approximate locations of the $F140M$, $F162M$, and $F182M$ JWST medium band filters which are discussed in \S\ref{sec:JWSTmedium}.}
\label{fig:LT_spectra}
\end{figure}

\subsection{Stellar Types M, L and T}
Brown dwarfs are sub-stellar objects distinguished from typical dwarfs
by their inability to maintain stable fusion of hydrogen, leaving
their cores supported by electron-degeneracy pressure. This arises
from their low masses ($<$0.08\,M$_\odot$) and results in low surface
temperatures (Luhman et al, 2012). They represent a continuation of
the mass/luminosity-temperature sequence for low mass stars beyond the
end of the main sequence, extending the range of stellar types from
mid-M to the later classes L, T and most recently Y.

Observationally, these sequences are defined by increasing prominence
of molecular absorption features in the atmospheres of the star, and
hence in the observed spectra. In particular molecular absorption in
CaH, FeH, CrH and H$_2$ are all possible at these low temperatures
(e.g. Gizis 1997, Burgasser et al 2003). As the temperature drops
(i.e. moving from M to L to T classes), the stellar spectrum becomes
redder, and molecular absorption features strengthen. This results
both in distinctive variation of narrowband indices, defined to map
individual line strengths, and also of broad-band photometric colours
(e.g. Burgasser et al 2003, 2007, Bochanski et al 2007).

In the context of high-redshift contamination, it is important to note that these low temperatures result
in sources that are intrinsically faint, and
spectra in which the continuum is interrupted and can appear to
show large breaks across narrow wavelength ranges, or in which the
flux can peak up above the noise level in narrow regions (see Fig.
1).

The classification of a sequence of stellar sub-types is based on the
observed evolution in the spectra of these stars with temperature, and
thus is defined and calibrated using a set of relatively bright,
nearby stars for which detailed properties can be measured. These are
then used as spectral standards against which other, more distant
stars can be calibrated (see e.g. Burgasser et al 2007). The
comprehensive SpeX Prism Spectral Library
\footnote{\url{http://pono.ucsd.edu/~adam/browndwarfs/spexprism/}} provides
a repository of data on these standards and other brown dwarfs.
Optical-Near-IR spectroscopy, spanning roughly 0.65-2.55$\,\mathrm{\mu m}$ and
obtained with the SpeX spectrograph on the NASA Infrared Telescope
Facility, are available from this source and show the effect of
temperature on the stellar spectrum. The spectral energy distributions
of L and T (L1$\to$T8) standards in this library are shown in Fig. 1,
with Table 2 providing the original data references for each star.

As can be seen, the strength and sharpness of the spectral absorption
breaks is increased towards later sub-types, with deep troughs and red
continua potentially leading to dropout colours in the near-infrared.

\subsection{Y-dwarf stars}
While the T spectral class extends to very low effective
temperatures and masses, it is not the end of the story. Just as the
new class `L' was defined in 1999 (Kirkpatrick et al 1999), and `T'
shortly thereafter (Burgasser et al 1999, Leggett et al 2000), so the
increasing availability of near-infrared data has recently allowed the
classification scheme to be extended to the coolest, lowest mass brown
dwarfs currently known: the Y class dwarfs. The first of these were
identified in near-infrared (3-10$\mu$m) data from the {\em Wide-field
Infrared Survey Explorer} ({\em WISE}) with WISEP J173835.52+273258.9
identified as a Y0 standard (Kirkpatrick et al 2012; Cushing et al
2011). They are characterized by the condensing out of various atomic
and molecular species as the temperature of the stellar atmosphere
drops, with inferred effective temperatures (based on spectral
modelling) below $\sim$600\,K and masses below
$\sim$20\,M$_\mathrm{Jup}$ (Cushing et al 2011).

Given the faint absolute magnitudes of these sources (M$_H>20$), their
red spectra and the difficulty of near-infrared spectroscopy, work on
defining standards and obtaining empirical SEDs for these sources is
continuing.  As a result, estimation of their impact on an
extragalactic survey must necessarily be based on modelled spectra,
checked against observations, rather than directly on observational
standards.

The brown dwarf atmosphere models of Morley et al (2012) explore the
impact of a variety of optically-thin and thick condensates forming
clouds in the atmospheres of T and Y dwarfs. They include opacities
for Cr, MnS and Na$_2$S amongst other species and span an effective
temperature range of 400\,K to 1300\,K, based on a combination of empirical data and theoretical predictions for the properties of these very low mass stars. They also probe two
possible surface gravities and the cases with no clouds and with a
range of condensate sedimentation efficiencies,
$f_\mathrm{sed}$. Comparison with observed late T-dwarfs suggest that
these models provide a better fit to observed spectra than earlier
attempts, although uncertainties remain (Morley et al 2012). In the
following we use these as templates and consider a range of models
from this data set that are believed to accurately reproduce the
spectra of Late-T and Y dwarfs. For completeness in the sections that
follow, we show all values of $f_\mathrm{sed}$ in the model grid,
while noting that the presence or absence of clouds has a larger
effect than the sedimentation efficiency for all but the coolest
stars.

\subsection{Surface densities}
Most existing surveys of distant galaxies dismiss the possibility that their samples
are heavily contaminated by brown dwarfs based on two criteria: the ability of {\em HST}
to separate galaxies from stars based on their spatial light profiles, and the low
inferred angular density of these sources on the sky.

As already noted, at faint magnitudes and as observations move into
near-infrared bands, the ability to reliably separate stars from
compact galaxies on morphological grounds becomes limited. Samples
which reject all unresolved sources risk underestimating the galaxy
population, while those which include them or reach to magnitudes
where the distinction is lost are liable to be contaminated.

So, assuming morphological information is unavailable, are the surface
densities of these sources on the sky sufficiently high to cause
concern? The honest answer, of course, is that it is hard to be
certain.  At faint magnitudes, we are probing brown dwarfs at greater
distances than those accessible to existing studies, and hence in
regions of the Galactic disc and halo that may differ significantly
than that in the Sun's immediate vicinity.  The possibility of
encountering regions overdense in brown dwarfs cannot be dismissed.
Nonetheless, making the crude assumption that the local region is, in
fact, typical of the disc population, we can gauge the scale of
potential contamination from these sources.

T-dwarfs have absolute magnitudes in the near-infrared M$_H\sim15-20$,
while Y-dwarfs typically have M$_H>20$. Thus deep surveys reaching to
an apparent magnitude m$_H=30$ (e.g. the Hubble Extreme Deep Field,
XDF, Illingworth et al. 2013) would be sensitive to late T and early Y
dwarfs out to distances of $\sim1.0$\,kpc. Kirkpatrick et al. (2012)
notes there are 33 known LTY dwarfs within 8\,pc of the Sun (3 L,
T$\ge22$, Y$\ge8$) suggesting a space density of
$\gtrsim$0.015\,pc$^{−3}$. This certainly presents a lower limit since
sources within well this volume are still being found and are not
included in this Fig. (e.g. Luhman 2013, Bihain et al 2013).

However, taking the Kirkpatrick et al sample as a basis, we would
expect $\sim2\times10^5$ sources out to a radius of 0.15\,kpc
(m$_H$=26) or $\sim$5.2\,deg$^{-2}$ - consistent with the surface density
of $z$-drop sources in UltraVISTA (Bowler et al. 2012). Extrapolating further,
out to a radius of 1\,kpc (m$_H$=30), we would expect
$\sim$1563\,deg$^{-2}$ or $\sim$0.4\,arcmin$^{-1}$.  Hence, one might
expect $\sim3-4$ such sources in a single WFC3 pointing such as the
XDF.  Given the relatively small number of $z-$ and $Y-$drop
candidates in such a pointing, this can represent a substantial
contamination fraction
 
Of course there are many uncertainties on this number - as
noted above, the catalogue of local LTY dwarfs used has already been shown to
be incomplete, and may also be unrepresentative of the population
along the sightline to any given deep field. Few fields on the sky are
as free from galactic extinction and angled as far from the Galactic
plane as the XDF. While the local measurements provide a density
within the plane, we have no constraint whatsoever on the population
of dwarf stars out of the Galactic plane or in the Galactic halo.
Inevitably, as deep fields probe larger areas, they will also probe
disc, or halo, substructure. Ultimately, the only way to be certain of
the impact of stellar contamination is to directly observe it.

\section{Photometric colours and potential contamination}\label{sec:photo}

\begin{figure*}
\centering
\includegraphics[width=40pc]{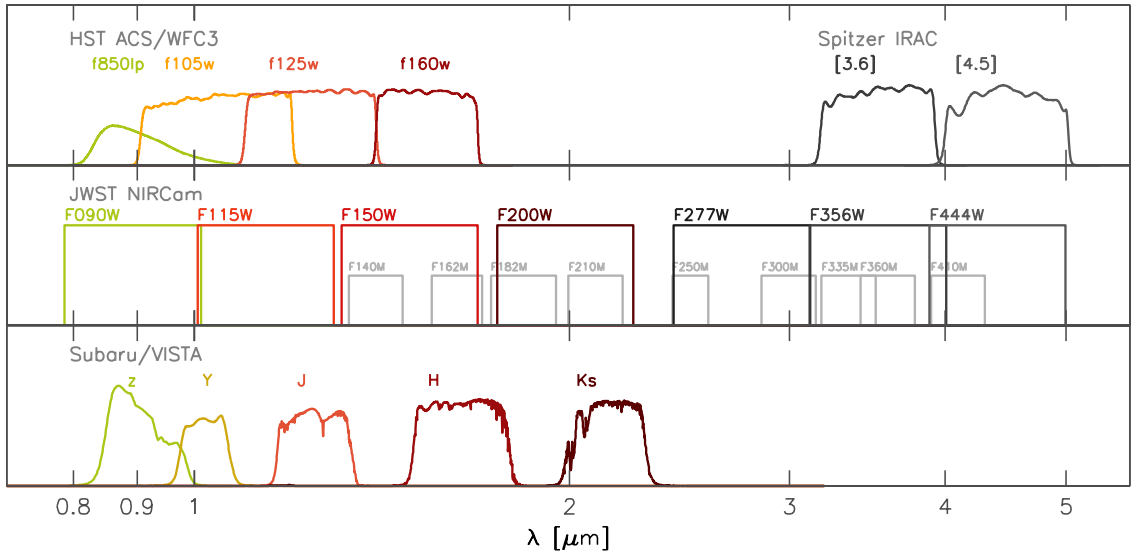}
\caption{The various filters considered in this study.}
\label{fig:filters}
\end{figure*}

In considering the potential for these sources to contaminate galaxy 
selections targeting the very distant ($z>7$) Universe, it is necessary first to consider the
disparate methods and data that are used to identify candidates. Key
to this is the question of which filters are used for the photometry
of a deep field. Much of this work to date has been based on {\em HST}
observations, which benefit both from lower backgrounds and from
higher angular resolution than is possible from the ground. However,
an increasing number of large fields have ground-based photometric
data reaching reasonable depths in the near-infrared - notably from the VISTA public
surveys. While these do not reach the depth of the most sensitive {\em
HST} surveys, they exceed the spatial coverage of such studies by
orders of magnitude, and thus are likely to be increasingly used to
select bright candidates for follow-up. In the future, this observational
effort will expand to include surveys undertaken with the {\em
James Web Space Telescope (JWST)}. As a larger telescope than {\em
HST}, this is likely to push still deeper, albeit still in relatively small
fields.

Strong emission features in the Earth's atmosphere render ground based
observations insensitive to some spectral regions, and photometric
filters are naturally designed to exclude noise due to this background
flux. Even in space, the bandwidth and spectral response functions of
filters are modified by detector sensitivity and changes in design
between instruments.  As Fig. \ref{fig:filters} illustrates, the
selection of filters available from the ground (with VISTA) and with
{\em HST} or {\em JWST} varies significantly.  While this has little
effect on the colours of smooth continuum sources, for those with
interrupted or line-dominated spectra (like dwarf stars or distant
galaxies) the choice of filter shape and central wavelength can have a
substantial effect on the measured fluxes and colours.

In this analysis we consider the potential that very-higih-redshift ($z\approx 6.5-9.0$) galaxy selections
are contaminated by cool dwarfs using the three filter combinations given in table 
\ref{tab:colourkey}. For {\em Hubble} and Subaru/VISTA filters we use the measured
filter transmission curves including detector and instrumentation transmission effects.
For {\em JWST} we simply consider the width and central wavelength of planned filters in
the near-infrared imaging camera, NIRCam. In Section \ref{sec:gt2} we extend this analysis to include observations at $\lambda>2\,\mathrm{\mu m}$.

\begin{table}
\caption{Key to the colours presented in Table \ref{tab:LTcolours}. \label{tab:colourkey}}
\centering
\begin{tabular}{lc}
\hline
\multicolumn{2}{c}{{\em Hubble}} \\
\hline
(1) & $z_{f850lp}-Y_{f105w}$\\
(2) & $Y_{f105w}-J_{f125w}$\\
(3) & $J_{f125w}-H_{f160w}$\\
\hline
\multicolumn{2}{c}{{\em JWST}} \\
\hline
(1) & $F090W-F115W$\\
(2) & $F115W-F150W$\\
(3) & $F150W-F200W$\\
\hline
\multicolumn{2}{c}{{\em Subaru} + {\em VISTA}} \\
\hline
(1) & $z-Y$\\
(2) & $Y-J$\\
(3) & $J-H$\\
\hline
\end{tabular}
\end{table}

We calculate the near-infrared colours expected for L and T spectral
standards (described in section \ref{sec:browndwarfs}) using these
filter transmission profiles, and present these in
table \ref{tab:LTcolours}. While all three filter combinations can
be classified as defining a $z-Y-J-H$ colour-colour space, with comparable
filter central wavelengths, the differences in colour can be several
tenths of a magnitude - significantly higher than the expected photometric
noise in deep surveys. 

\begin{table*}
\caption{The near-infrared colours of L and T spectral standards computed using spectra from the SpeX Prism Spectral Library using the {\em HST} filter set. The key for the various colours is shown in Table \ref{tab:colourkey}. \label{tab:LTcolours}}
\centering
\begin{tabular}{lccccccccc}
\hline
Spectral Type & \multicolumn{3}{c}{{\em Hubble}} & \multicolumn{3}{c}{{\em JWST}} & \multicolumn{3}{c}{{\em Subaru} + {\em VISTA}} \\
\hline
 & (1) & (2) & (3) & (1) & (2) & (3) & (1) & (2) & (3)  \\
L1 & 1.03 & 0.52 & 0.18 & 1.36 & 0.31 & 0.11 & 0.79 & 0.85 & 0.26  \\
L3 & 1.14 & 0.65 & 0.29 & 1.56 & 0.43 & 0.25 & 0.87 & 1.07 & 0.39  \\
L4 & 1.15 & 0.64 & 0.26 & 1.58 & 0.41 & 0.29 & 0.86 & 1.06 & 0.38  \\
L4.5 & 1.15 & 0.63 & 0.39 & 1.59 & 0.5 & 0.3 & 0.91 & 1.06 & 0.49  \\
L6 & 1.12 & 0.7 & 0.55 & 1.59 & 0.69 & 0.48 & 0.94 & 1.07 & 0.68  \\
L8 & 1.1 & 0.63 & 0.51 & 1.55 & 0.61 & 0.42 & 1.0 & 0.91 & 0.61  \\
L9 & 1.14 & 0.58 & 0.49 & 1.61 & 0.49 & 0.2 & 1.06 & 0.9 & 0.53  \\
\hline
T0 & 1.13 & 0.56 & 0.31 & 1.62 & 0.3 & 0.14 & 1.08 & 0.88 & 0.32  \\
T1 & 1.21 & 0.54 & 0.33 & 1.69 & 0.28 & 0.02 & 1.18 & 0.83 & 0.33  \\
T2 & 1.38 & 0.59 & 0.23 & 1.93 & 0.15 & 0.03 & 1.47 & 0.91 & 0.19  \\
T3 & 1.39 & 0.56 & 0.08 & 1.91 & -0.05 & -0.16 & 1.53 & 0.85 & -0.03  \\
T4 & 1.57 & 0.66 & -0.07 & 2.18 & -0.16 & -0.18 & 1.71 & 1.01 & -0.23  \\
T5 & 1.52 & 0.63 & -0.37 & 2.2 & -0.58 & -0.35 & 1.85 & 0.9 & -0.65  \\
T6 & 1.52 & 0.71 & -0.43 & 2.34 & -0.6 & -0.37 & 1.96 & 0.95 & -0.74  \\
T7 & 1.66 & 0.7 & -0.43 & 2.57 & -0.74 & -0.22 & 2.33 & 0.84 & -0.83  \\
T8 & 1.62 & 0.74 & -0.32 & 2.55 & -0.69 & -0.49 & 2.43 & 0.84 & -0.78  \\
\hline
\end{tabular}
\end{table*}

\begin{figure}
\centering
\includegraphics[width=20pc]{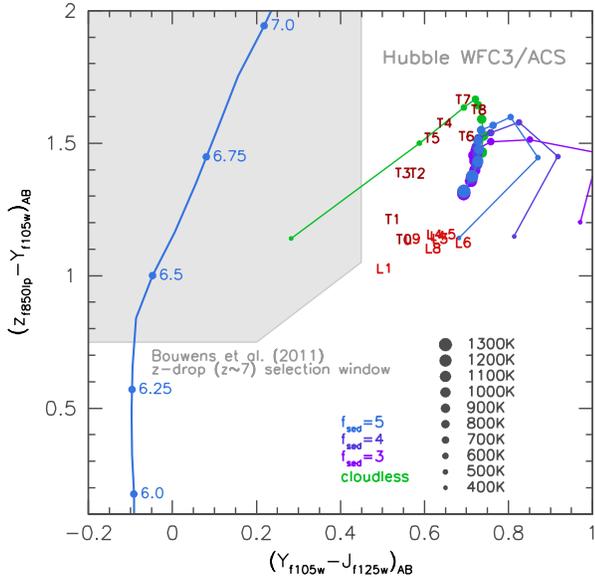}
\includegraphics[width=20pc]{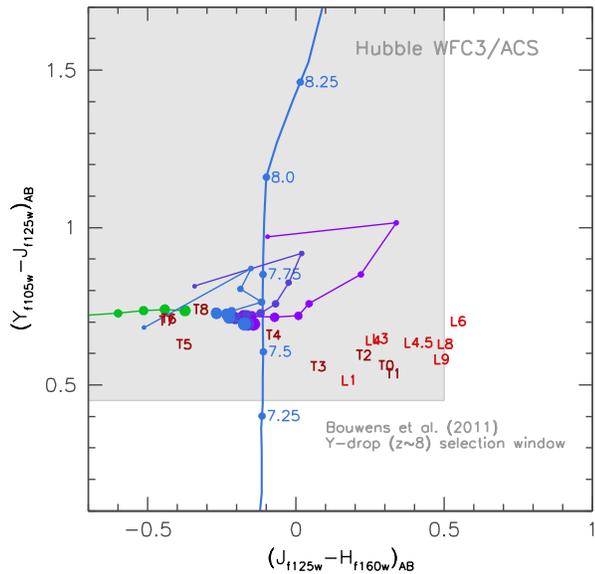}
\caption{{\em Hubble} ACS/WFC3 colour-colour diagrams showing the location of L and T dwarf spectral standards compared to the Bouwens et al. (2011) selection windows and the expected redshift track of a high-redshift dust-free star-forming galaxy. Model Y dwarfs are shown by connected tracks - see upper panel for key.}
\label{fig:obs_HST}
\end{figure}

\begin{figure}
\centering
\includegraphics[width=20pc]{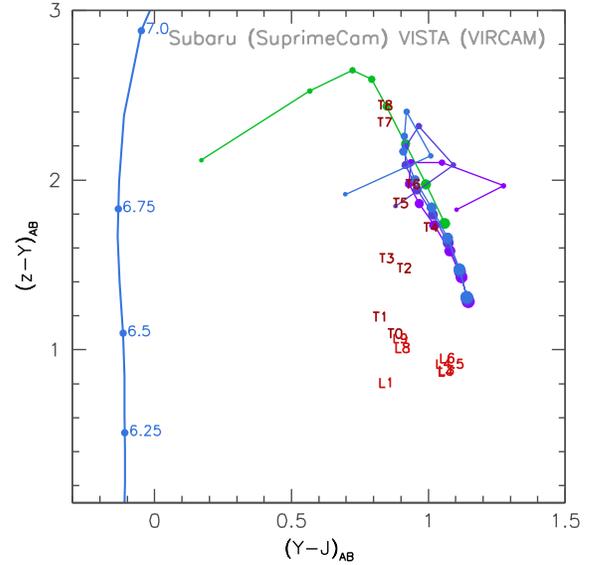}
\includegraphics[width=20pc]{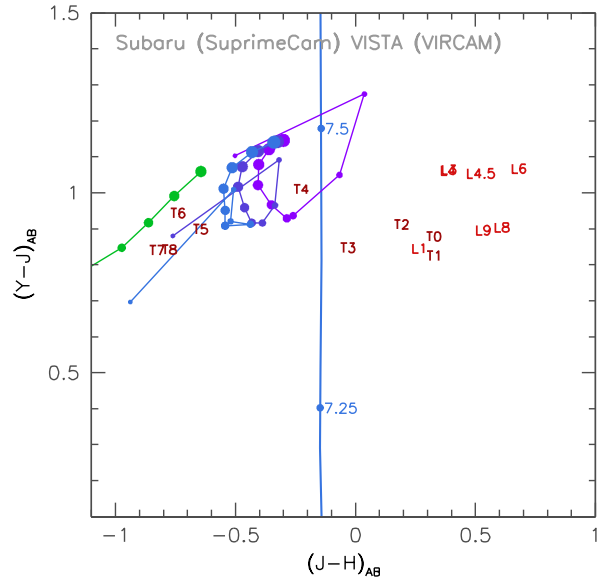}
\caption{Subaru SuprimeCam and VISTA VIRCam colour-colour diagrams, with the location of low-mass stars, and the tracks of high-redshift galaxies as in fig. \ref{fig:obs_HST}.}
\label{fig:obs_VISTA}
\end{figure}

\begin{figure}
\centering
\includegraphics[width=20pc]{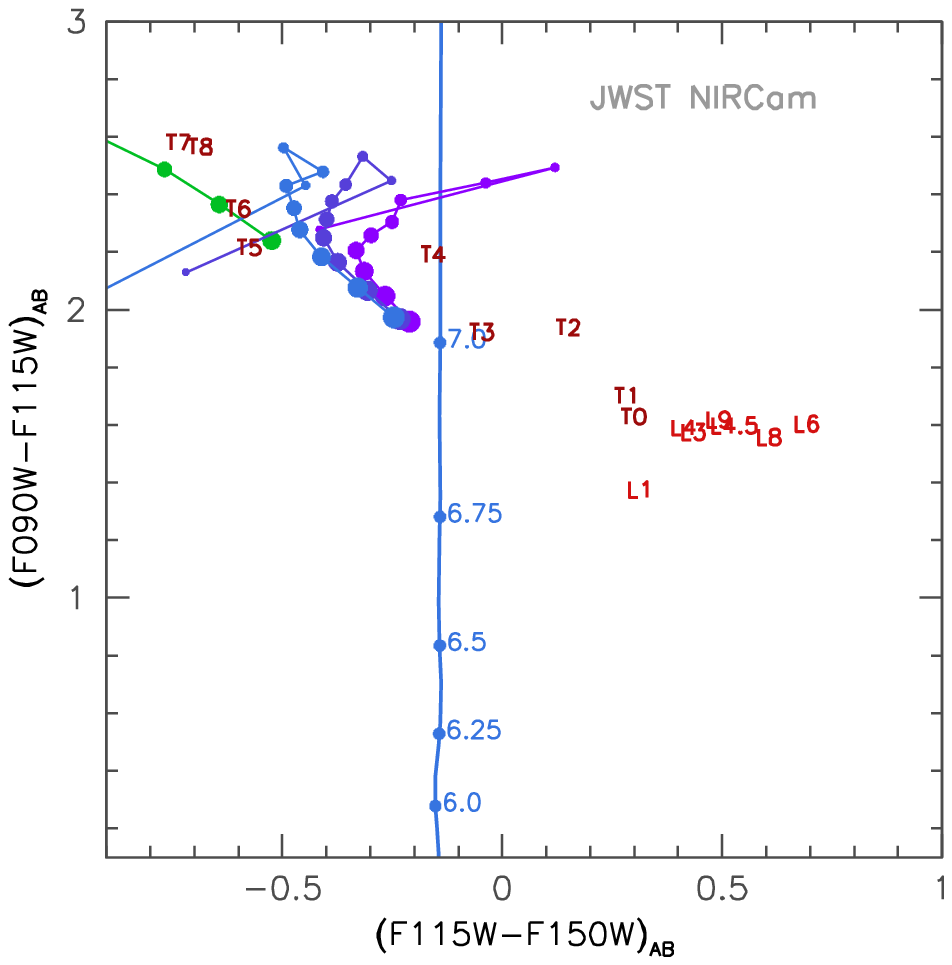}
\includegraphics[width=20pc]{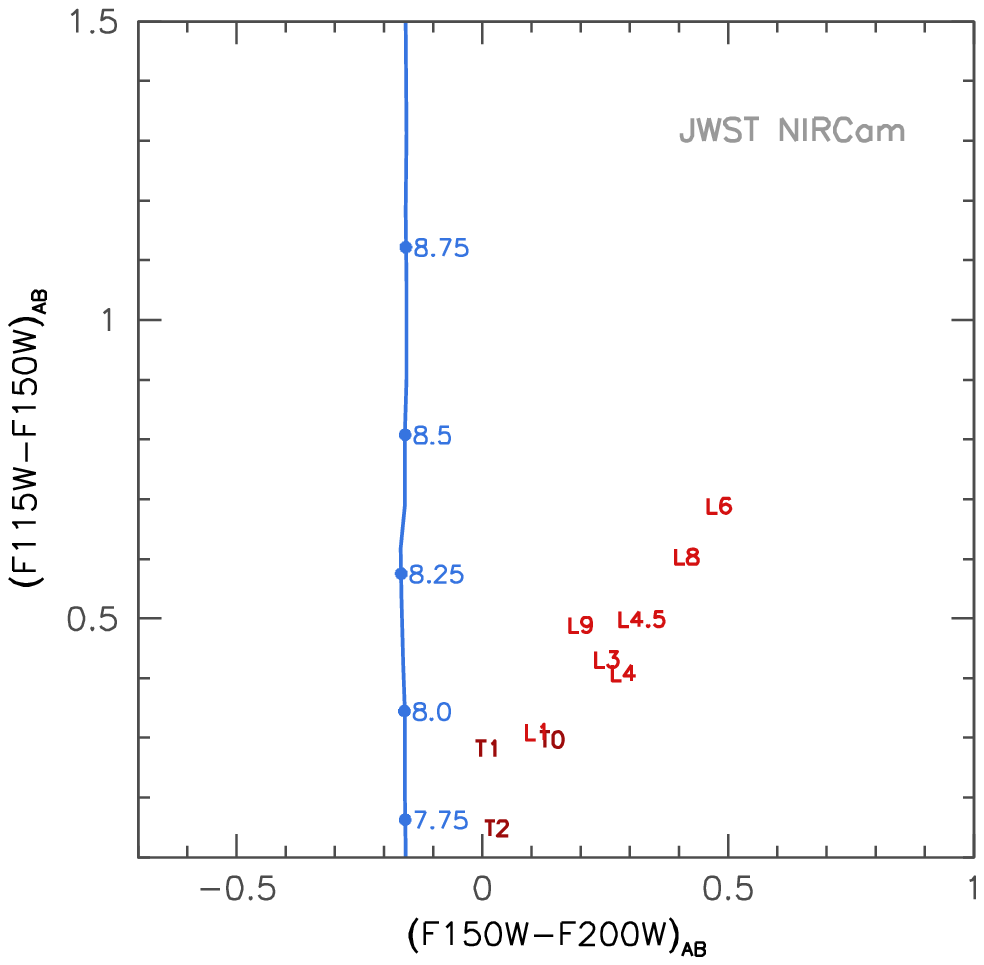}
\caption{{\em JWST} NIRCam colour-colour diagrams, with the location of low-mass stars, and the tracks of high-redshift galaxies as in fig. \ref{fig:obs_HST}. In the lower panel, the model Y dwarfs lie off the plot ($F115W-F150W<0.1$).}
\label{fig:obs_JWST}
\end{figure}

We illustrate this graphically for {\em Hubble}, Subaru/VISTA and {\em
JWST} filters respectively in
figures \ref{fig:obs_HST}, \ref{fig:obs_VISTA} and \ref{fig:obs_JWST}.
In Fig. \ref{fig:obs_HST} we also show two of the most commonly used
photometric selection criteria used to identify distant galaxy
candidates.  The {\em HST} criteria used to select $z$-drop ($z>6.5$)
galaxies \citep[e.g.][]{2011ApJ...737...90B} are relatively
clean. While the L and T-dwarf standards have colours within 0.05
magnitudes of the colour-colour boundary for selection, this boundary
itself lies significantly redwards of the expected galaxy
colours. Dust reddening in distant galaxies will work to reduce this
separation, but is unlikely to account for a 0.4 magnitude shift in
colour across a relatively small rest-frame wavelength interval. It is
clear that any low signal-to-noise measurements, or very red
candidates (in $Y_{f105w}-J_{f125w}$) need to be approached with
caution, but for a `normal' candidate, the ambiguity is likely to be
small.

By contrast the commonly applied criteria used to select $Y$-drop
($z\sim7.5$) galaxies \citep[again see][]{2011ApJ...737...90B} is
problematic. The colour-selection space is populated by a large range
of L through T-dwarf stars. Contamination by cool stars is likely to be
limited by only two factors: the density of such sources on the sky at
the depths under consideration, and the degree to which $z\sim8$
galaxies can be reliably distinguished from point sources at these
depths. For the brightest objects, it may also be possible to use the $z$-band as a discriminant, as cool stars are expected to have $z-Y$ colours in the range $1-2$ ({\em Hubble}/WFC3) and $0.8-2.5$, while galaxies at $z>7.5$ are expected to be virtually undetected.

Studies using {\em JWST} and Subaru/VISTA photometric filters are yet
to be published, so no well established colour selection criteria for
distant galaxies exist in these filter sets - one motivation for this
paper is to advise on such selections. 

Comparison between Fig. \ref{fig:obs_VISTA} and
Fig. \ref{fig:obs_HST} suggests that similar issues will affect the
ground-based samples to those currently observed from
space. Contamination in ground based samples is more of a problem
since stars cannot be reliably excluded on morphological grounds, even
at relatively bright magnitudes. The locus occupied by $z-$drop
galaxies at $z\sim6.5-7.0$ is clearly distinct from that expected for
L and T dwarf stars (separated by $\sim$0.9 magnitudes), while that
for $Y-$drop ($z\sim7.5$) sources is likely to be heavily
contaminated.  It may be possible to obtain a cleaner selection at
$z>7.5$ by moving to a higher $Y-J$ colour-colour cut, but doing so is
likely to require extremely deep $Y$ band data, and so requires careful
consideration at the survey design stage. Either way, ground-based
selection of galaxies at $z\sim7.0-7.5$ is likely to be either
incomplete or contaminated, and this should be taken into account when
properties of this population are considered.

As Fig. \ref{fig:obs_JWST}
demonstrates, the use of {\em JWST} NIRCam filters to select $z\sim7$
galaxies is likely to be challenging due to contamination by mid-T
dwarf stars. The equivalent of a $z$-drop selection will not be straightforwardly
possible. By contrast, selection at $z<6.5$ and $z>8.5$ are likely
to be relatively free of L and T dwarf contamination. It is
interesting to note that cool stars also seem to form a well-defined
sequence that diverges from that of galaxies in the $F115W-F150W$ vs
$F150W-F200W$ colour-colour plane.  Given sufficiently precise
photometry it may be possible to exclude some of this locus with well
selected colour-colour cuts, allowing selection of $z\sim8.25$
galaxies.

However, as discussed in section \ref{sec:browndwarfs}, the now well-defined L and T dwarf star populations do not tell the complete story here. Spectral standards for the new Y dwarf population are still being established. The faint magnitudes of those already identified and limitations of instrumentation mean that high-signal-to-noise spectroscopy for these stars across a long wavelength baseline is not available in the same way as it is for L and T dwarfs.

In tables \ref{tab:modelcolours_HST}, \ref{tab:modelcolours_VISTA}
and \ref{tab:modelcolours_JWST} we calculate the predicted colours for
these very cool sources using the spectral synthesis models
of \citet{2012ApJ...756..172M}, described in section \ref{sec:browndwarfs}. In
figures \ref{fig:obs_HST}, \ref{fig:obs_VISTA}
and \ref{fig:obs_JWST} we present the calculated colours
graphically, as for the L and T dwarfs above. Each set of models is
shown as a connected track linking models with the same cloud
composition assumptions, but with different temperatures. In each
case, the lowest temperature models ($\sim$400K) and those models
which neglect the presence of molecular clouds in the stellar
atmosphere produce the most extreme colours. Uncertainties in our
current limited understanding of Y dwarfs therefore makes design of
colour selection criteria to exclude these sources challenging.

Nonetheless, it is clear that the Y dwarf population is likely to
occupy a similar colour-colour parameter space to L and T dwarfs -
likely contaminating {\em HST} or Subaru/VISTA $Y$-drop samples, and
presenting a challenge for {\em JWST} NIRCam $z$-drop selections. Y dwarfs are likely to be redder in $Y-J$ than T dwarfs by 
up to 0.3 magnitudes for the coolest sources - perhaps challenging high
redshift identifications that would otherwise by considered `unambiguous'. It
is worth noting too that while current Y dwarf observations favour
$f_{sed}\sim4-5$, they do not exclude cloudless models. If future
observations favour these, the potential for contamination in {\em
HST} or Subaru/VISTA $z$-drop samples also rises beyond that currently
estimated, with the coolest dwarf stars entering the currently used
colour-selection windows.

Alternate colour selection criteria for distant galaxy samples using
the {\em HST}/WFC3 filter set have been suggested by
\citet{2011MNRAS.414.1455L,2013MNRAS.429..150L} and \citet{2011MNRAS.411...23W}.
We note that the \citet{2011MNRAS.411...23W} colour selection offers little advantage
over that of \citet{2011ApJ...737...90B} for the exclusion of cool
stars from $z$-drop samples and may indeed admit some T and Y dwarfs that
lie clear of the \citet{2011ApJ...737...90B} selection, due to
photometric scatter.  The \citet{2011MNRAS.414.1455L} criteria for
selection of $Y$-drops are significantly redder in $Y-J$ than those
of \citet{2011ApJ...737...90B}. This does serve to reduce T dwarf
contamination, and will likely be effective against most Y dwarfs,
although we note that some of the coolest stellar models still satisfy
these criteria, even before photometric scatter is taken into account.

\subsection{The utility of near-infrared observations at $\lambda > 2\,\mathrm{\mu m}$}\label{sec:gt2}

Given the difficulty in cleanly separating low-mass stars from high-redshift star-forming galaxies using observations at $\lambda < 2\,\mathrm{\mu m}$ it is useful to investigate whether the addition of near-IR observations at $\lambda > 2\,\mathrm{\mu m}$ is of benefit. Deep near-IR observations at $\lambda>2\,\mathrm{\mu m}$ are currently available from the {\em Spitzer Space Observatory} using the Infrared Array Camera (IRAC) and in the future from {\em JWST}. While the recent {\em Wide Field Infrared Survey Explorer (WISE)} provides imaging at $3.4$, $4.6$, $12$ and $22\,\mu m$ this is not well matched, in terms of sensitivity, to either the deep {\em Hubble} or ground-based searches for very-higih-redshift galaxies.  

\begin{figure}
\centering
\includegraphics[width=20pc]{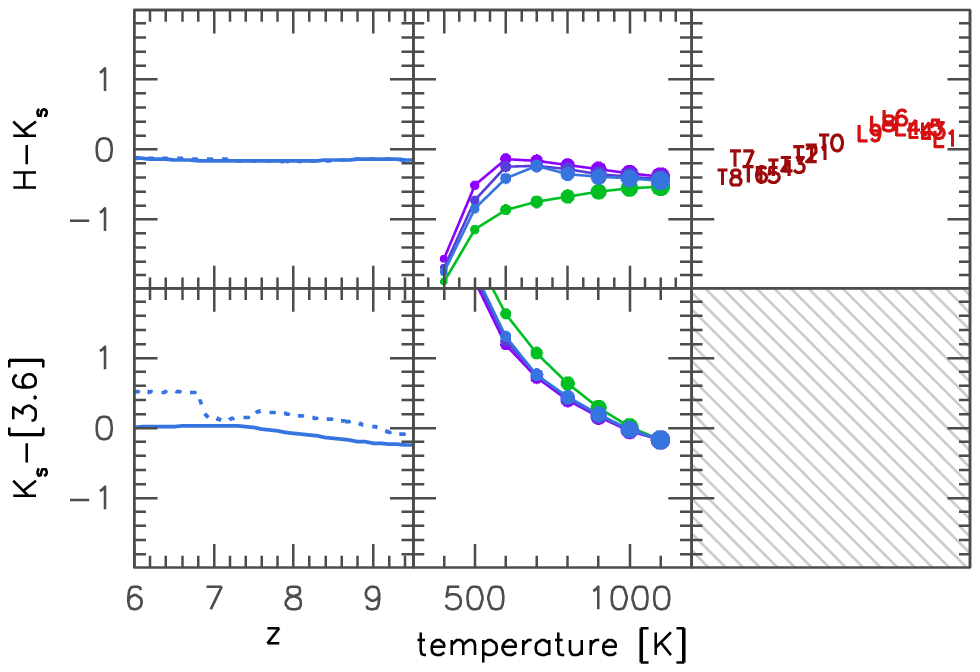}
\includegraphics[width=20pc]{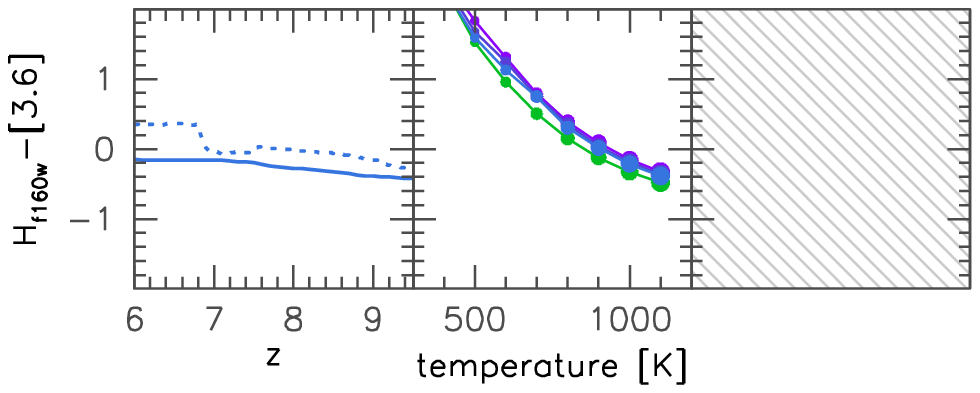}
\includegraphics[width=20pc]{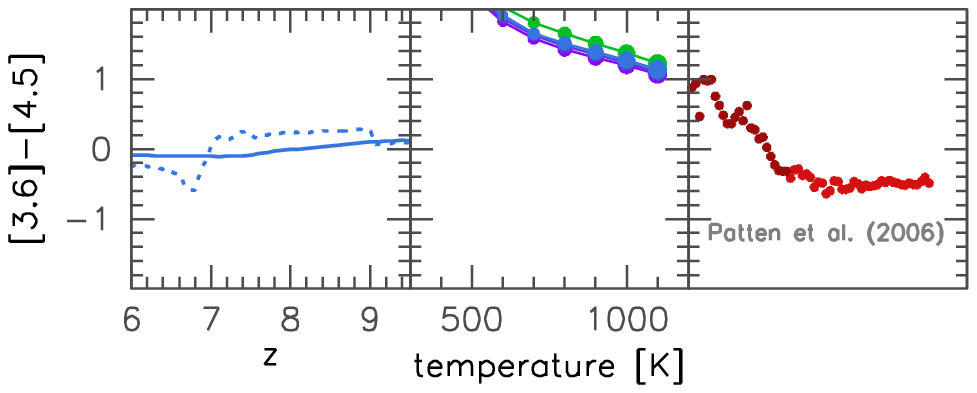}
\caption{VISTA/VIRCam, {\em HST}/WFC3, {\em Spitzer}/IRAC colours of high-redshift star-forming galaxies ({\em left}, as a function of redshift), the spectral synthesis models of \citet{2012ApJ...756..172M} ({\em middle}), and observed L and T stars ({\em right}). The solid blue curve shows the pure-stellar colour of a star-forming galaxy predicted from the {\sc Pegase.2} stellar population synthesis model assuming $100\,{\rm Myr}$ previous duration of constant star formation and $Z=0.0004$. The dashed curve shows the colour when self-consistent nebular emission is included. Near-IR spectroscopy of the L and T standards included in the SpeX library does not extend to cover the {\em Spitzer}/IRAC. In their place we plot the observed colours from \citet{2006ApJ...651..502P} (corrected to the AB magnitude system).}
\label{fig:obs_HSTVISTASpitzer_sc}
\end{figure}

In Fig. \ref{fig:obs_HSTVISTASpitzer_sc} we show the $H-K_{s}$, $K_{s}-$[3.6], $H_{f160w}-$[3.6] and [3.6]$-$[4.5] colours of high-redshift star-forming galaxies, the spectral synthesis models of \citet{2012ApJ...756..172M}, and the observed L and T standards from the SpeX library. Unfortunately, near-IR spectroscopy at $>2.5\,\mathrm{\mu m}$ (i.e. covering the IRAC channels) is unavailable for the L and T standards in the SpeX library. However, for the [3.6]$-$[4.5] colour we augment our analysis using the direct observations of \citet{2006ApJ...651..502P}. 

The $H-K_{s}$ colours of both the models and the L and T standards are similar to those expected for high-redshift galaxies. While it is not possible to determine the $H_{f160w}-$[3.6] colour for the L and T standards the colours of the hottest \citet{2012ApJ...756..172M} models (which are similar to the observed late T standards) overlap with the expected colours of very-higih-redshift star-forming galaxies. A similar result is also obtained for the using the {\em VISTA}/VIRCam $K_{s}$-band observations. While the [3.6]$-$[4.5] colours of the \citet{2012ApJ...756..172M} models remain consistently redder that those expected for high-redshift galaxies, the L and early T standards are much bluer and overlap those of very-higih-redshift sources. It is also worth noting that at very-higih-redshift the IRAC [3.6] and [4.5] channels probe the rest-frame optical and are expected to be strongly affected by nebular emission \citep[e.g.][]{2013MNRAS.435.2885W} which can result in much redder expected colours for the high-redshift galaxies. 

\begin{figure}
\centering
\includegraphics[width=20pc]{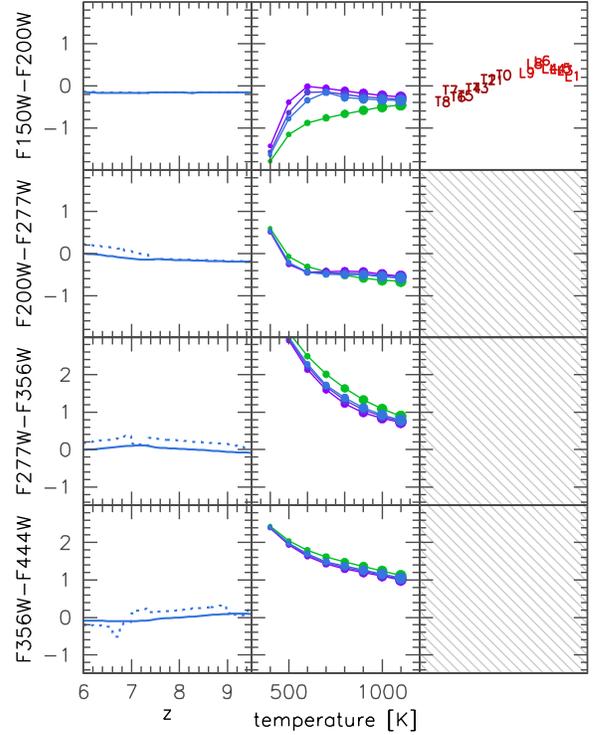}
\caption{The same as Fig. \ref{fig:obs_HSTVISTASpitzer_sc} but showing the result for {\em JWST}/NIRCam colours.}
\label{fig:obs_JWST_sc}
\end{figure}

In the future NIRCam on {\em JWST} will also be capable of obtaining observations beyond $2,\mathrm{\mu m}$ with sensitivities\footnote{Version P1.6 of the {\em JWST} Prototype Exposure Time Calculator (ETC, \url{http://jwstetc.stsci.edu/etc/}) suggests a peak S/N for a given exposure time and flat $f_{\nu}$ spectrum is achieved using the $F277W$ ($\lambda_{c}=2.77\,{\rm\mu m}$) filter with observations in $F356W$ ($\lambda_{c}=3.56\,{\rm\mu m}$) reaching similar sensitivities to those utilising $F200W$ ($\lambda_{c}=2.0\,{\rm\mu m}$).} comparable with those at $<2\,\mathrm{\mu m}$. Fig. \ref{fig:obs_JWST_sc} shows the expected colours, using the {\em JWST}/NIRCam wide set of filters, of high-redshift star-forming galaxies, observed L and T stars, and the spectral synthesis models of \citet{2012ApJ...756..172M}. Neither the $F150W-F200W$ or $F200W-F277W$ colours offer the ability to discriminate between low-mass stars and high-redshift galaxies. For both the $F277W-F356W$ and $F356W-F444W$ the predicted colours of the \citet{2012ApJ...756..172M} models are somewhat redder than that expected for dust-free high-redshift star-forming galaxies. However, extrapolating the temperature trend, suggests the L and T classes will have colours closer to those expected for high-redshift galaxies. Further, dust and nebular emission will both redden the expected $F277W-F356W$ and $F356W-F444W$ colours of high-redshift objects.

\begin{table*}
\caption{The near-infrared colours of model low-mass stars assuming the {\em Hubble} ACS and WFC3 filter set. The key for the various colours is given in Table \ref{tab:colourkey}.\label{tab:modelcolours_HST}}
\centering
\begin{tabular}{lcccccccccccc}
             & \multicolumn{3}{c}{cloud free} & \multicolumn{3}{c}{$f_{sed}=3$} & \multicolumn{3}{c}{$f_{sed}=4$} & \multicolumn{3}{c}{$f_{sed}=5$} \\
\hline
 Temperature & (1) & (2) & (3) & (1) & (2) & (3) & (1) & (2) & (3) & (1) & (2) & (3) \\
\hline
400 & 1.14 & 0.28 & -1.03 & 1.2 & 0.97 & -0.1 & 1.15 & 0.81 & -0.34 & 1.14 & 0.68 & -0.51  \\
500 & 1.5 & 0.59 & -1.05 & 1.46 & 1.02 & 0.34 & 1.45 & 0.92 & 0.02 & 1.45 & 0.87 & -0.15  \\
600 & 1.63 & 0.69 & -0.88 & 1.51 & 0.85 & 0.22 & 1.58 & 0.83 & -0.02 & 1.6 & 0.81 & -0.19  \\
700 & 1.67 & 0.72 & -0.72 & 1.51 & 0.76 & 0.04 & 1.54 & 0.76 & -0.07 & 1.57 & 0.76 & -0.12  \\
800 & 1.64 & 0.73 & -0.6 & 1.47 & 0.72 & 0.01 & 1.52 & 0.73 & -0.12 & 1.55 & 0.73 & -0.22  \\
900 & 1.59 & 0.74 & -0.51 & 1.45 & 0.71 & -0.07 & 1.49 & 0.72 & -0.17 & 1.51 & 0.73 & -0.24  \\
1000 & 1.53 & 0.74 & -0.44 & 1.43 & 0.72 & -0.15 & 1.46 & 0.73 & -0.22 & 1.48 & 0.73 & -0.27  \\
1100 & 1.47 & 0.74 & -0.37 & 1.4 & 0.72 & -0.18 & 1.42 & 0.72 & -0.23 & 1.43 & 0.73 & -0.27  \\
\hline
\end{tabular}
\end{table*}

\begin{table*}
\caption{The near-infrared colours of model low-mass stars assuming the Subaru SuprimeCam and VISTA VIRCAM filter set. The key for the various colours is given in Table \ref{tab:colourkey}.\label{tab:modelcolours_VISTA}}
\centering
\begin{tabular}{lcccccccccccc}
             & \multicolumn{3}{c}{cloud free} & \multicolumn{3}{c}{$f_{sed}=3$} & \multicolumn{3}{c}{$f_{sed}=4$} & \multicolumn{3}{c}{$f_{sed}=5$} \\
\hline
 Temperature & (1) & (2) & (3) & (1) & (2) & (3) & (1) & (2) & (3) & (1) & (2) & (3) \\
\hline
400 & 2.12 & 0.17 & -1.47 & 1.83 & 1.1 & -0.5 & 1.85 & 0.88 & -0.76 & 1.92 & 0.7 & -0.94  \\
500 & 2.52 & 0.57 & -1.46 & 1.97 & 1.27 & 0.04 & 2.09 & 1.09 & -0.32 & 2.14 & 1.01 & -0.51  \\
600 & 2.65 & 0.72 & -1.28 & 2.1 & 1.05 & -0.07 & 2.32 & 0.97 & -0.34 & 2.4 & 0.92 & -0.52  \\
700 & 2.59 & 0.79 & -1.11 & 2.1 & 0.94 & -0.26 & 2.2 & 0.92 & -0.39 & 2.26 & 0.91 & -0.44  \\
800 & 2.43 & 0.85 & -0.97 & 1.98 & 0.93 & -0.29 & 2.09 & 0.92 & -0.43 & 2.17 & 0.91 & -0.54  \\
900 & 2.21 & 0.92 & -0.86 & 1.86 & 0.97 & -0.35 & 1.94 & 0.96 & -0.46 & 2.0 & 0.95 & -0.54  \\
1000 & 1.97 & 0.99 & -0.75 & 1.73 & 1.02 & -0.41 & 1.8 & 1.02 & -0.49 & 1.84 & 1.01 & -0.55  \\
1100 & 1.74 & 1.06 & -0.64 & 1.58 & 1.08 & -0.4 & 1.63 & 1.07 & -0.47 & 1.66 & 1.07 & -0.51  \\
\hline

\end{tabular}
\end{table*}

\begin{table*}
\caption{The near-infrared colours of model low-mass stars assuming the {\em JWST} NIRCam filter set. The key for the various colours is given in Table \ref{tab:colourkey}.\label{tab:modelcolours_JWST}}
\centering
\begin{tabular}{lcccccccccccc}
             & \multicolumn{3}{c}{cloud free} & \multicolumn{3}{c}{$f_{sed}=3$} & \multicolumn{3}{c}{$f_{sed}=4$} & \multicolumn{3}{c}{$f_{sed}=5$} \\
\hline
 Temperature & (1) & (2) & (3) & (1) & (2) & (3) & (1) & (2) & (3) & (1) & (2) & (3) \\
\hline
400 & 1.79 & -1.64 & -2.04 & 2.28 & -0.42 & -1.74 & 2.13 & -0.72 & -1.86 & 2.04 & -0.94 & -1.92  \\
500 & 2.37 & -1.47 & -1.77 & 2.49 & 0.12 & -1.08 & 2.45 & -0.25 & -1.3 & 2.43 & -0.45 & -1.43  \\
600 & 2.59 & -1.24 & -1.63 & 2.44 & -0.04 & -0.84 & 2.53 & -0.32 & -0.97 & 2.56 & -0.5 & -1.14  \\
700 & 2.63 & -1.05 & -1.57 & 2.38 & -0.23 & -0.89 & 2.43 & -0.36 & -0.99 & 2.48 & -0.41 & -1.0  \\
800 & 2.58 & -0.9 & -1.51 & 2.3 & -0.25 & -0.93 & 2.38 & -0.39 & -1.02 & 2.43 & -0.49 & -1.11  \\
900 & 2.49 & -0.77 & -1.43 & 2.26 & -0.3 & -0.96 & 2.31 & -0.4 & -1.06 & 2.35 & -0.47 & -1.12  \\
1000 & 2.36 & -0.64 & -1.31 & 2.21 & -0.33 & -0.97 & 2.25 & -0.41 & -1.04 & 2.28 & -0.46 & -1.09  \\
1100 & 2.24 & -0.52 & -1.19 & 2.13 & -0.31 & -0.94 & 2.16 & -0.37 & -1.0 & 2.18 & -0.41 & -1.04  \\
\hline

\end{tabular}
\end{table*}

\subsection{{\em JWST} medium band observations}\label{sec:JWSTmedium}

\begin{figure}
\centering
\includegraphics[width=20pc]{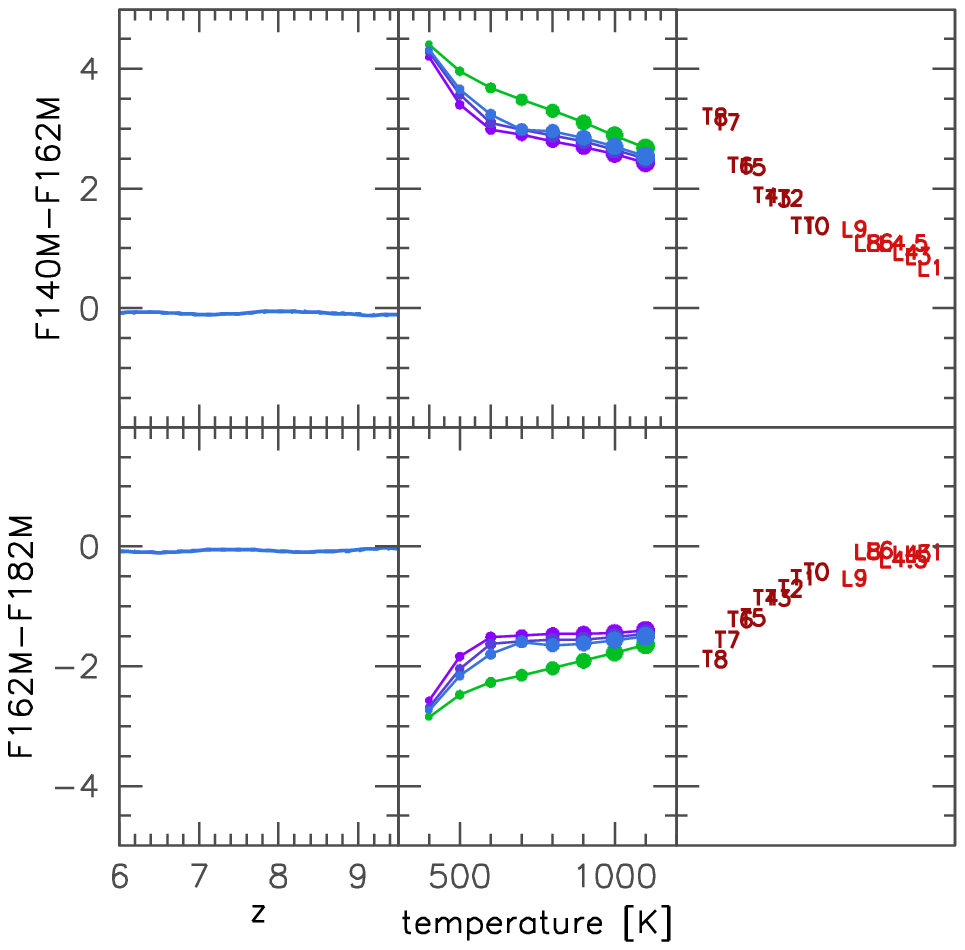}
\caption{The same as Fig. \ref{fig:obs_HSTVISTASpitzer_sc} but showing the result for {\em JWST}/NIRCam medium-band colours. The prediction including the effect of nebular emission is almost identical to that without because of the lack of strong emission lines falling in any of the filters at these redshifts.}
\label{fig:obs_JWST_sc_m}
\end{figure}

NIRCam on {\em JWST} will also be equipped with several medium band filters (shown in Fig. \ref{fig:filters}). These can be used to exploit the near-IR $\lambda<2\,\mathrm{\mu m}$ spectral energy distributions of late L and T-dwarfs which are punctuated by strong H$_2$O and CH$_4$ bands. As can be seen clearly in Fig. \ref{fig:LT_spectra}, the $F140M$ and $F182M$ filters coincide with deep molecular absorption bands, while the $F162M$ filter probes a peak in the SED. This results in very red $F140M-F162M$ and very blue $F162M-F182M$ colours for the late L and T-dwarfs, as can be seen in Fig. \ref{fig:obs_JWST_sc_m}. At the same time, the $F140M-F162M$ and $F162M-F182M$ colours are expected to be close to zero for high-redshift star-forming galaxies. While the L-dwarf population has $F140M-F162M$ and $F162M-F182M$ colours similar to those expected for high-redshift galaxies these can be separated from those at $z\sim 7$ using the $F115W-F150W$ colour as shown in Fig. \ref{fig:obs_JWST}. Hence the addition of imaging in selected JWST medium band filters could greatly enhance the reliability and utility of higih-redshift galaxy surveys in the future.

\section{Potential for Contamination of Spectroscopy}\label{sec:spectro}

Given that Y class dwarf stars are liable to contaminate photometric samples -  at a low but non-zero level - it is informative to consider whether spectroscopy can reliably identify these sources.
The vast majority of proposed $z>7$ candidate dropout galaxies are too faint for spectroscopy, even with 8\,m class telescopes. Nonetheless, a fraction of the brighter targets (for example, those from the CANDELS GOODS-S field) have now been subject to deep spectroscopic follow up and an even smaller fraction of these confirmed \citep[][]{2012MNRAS.427.3055C, caruana,2011ApJ...743..132P,2012ApJ...744...83O}. 

Given the faint near-infrared photometry of these sources, even the deepest spectroscopy yet undertaken does not reach the continuum level. As a result, these confirmations are entirely dependent on the detection of a single emission line appearing above the noise level in the spectrum, and its identification as the Lyman-$\alpha$ emission line of Hydrogen. Not only is a single emission line vulnerable to redshift misinterpretation (for example, the close [OII]\,3727\AA\ doublet, H$\alpha$\,6563\AA\ and the [OIII]\,5007\AA\ line are all relatively isolated and could appear as single line detections at low signal-to-noise), but it is also more prone to distortion by the noise characteristics of the data than a multi-line detection.

The peril of working at the low signal-to-noise demanded by these faint targets is illustrated by the $z\sim8.6$ identification proposed by \citet{2010Natur.467..940L} for Y-drop UDFy-38135539, based on a single line observed in deep SINFONI data. Later investigation failed to verify the existence of this line, suggesting it may have been a rare noise fluctuation \citep{2013MNRAS.430.3314B}. A second detection of line emission - a $z$-drop initially observed by \citet{2010ApJ...725L.205F} has proved equally difficult to confirm in deeper observations \citep{2012MNRAS.427.3055C,caruana}. This may equally be a result of noise in the original data, or potentially a noise fluctuation overlying a genuine emission line feature in the \citet{2010ApJ...725L.205F} observations and boosting it past a detection threshold \citep[][find a 3\,$\sigma$ blip in their deeper spectrum at the same wavelength]{caruana}.  Certainly, working at the typical `detection' signal-to-noise thresholds of $\sim5-10$ can lead to substantial variation in the flux value measured, due to random noise.

While interpretation of any given line can be challenging, one assumption generally holds: that the detection of line emission likely rules out identification as a cool dwarf star in our own galaxy. The peaky, interrupted spectra of the coolest dwarf stars make this a far from safe assumption. A simple question remains unanswered: assuming that a photometrically selected $z\sim7-9$ candidate source is followed up using a typical spectroscopic strategy, and that emission spatially coincident with it is detected, occupying a narrow wavelength range, to what degree can that `line emission' rule out identification as a late-T or Y-class dwarf star?

In detail, of course, this is an unanswerable question.  As discussed earlier (section \ref{sec:browndwarfs}), the spectra of modelled low mass stars are highly sensitive to the assumptions involved in the modelling. Chemical composition and the assumed physics of cloud formation in the molecular gas atmospheres of these dwarf objects can have substantial impact on the predicted spectroscopic characteristics.  Observations of Y-dwarfs at deep-field depths are as challenging and immature as the higih-redshift observations we seek to distinguish them from, but we can nonetheless explore their likely spectroscopic characteristics using a mixture of modelling and the relatively bright `template' sources observed in the Sun's local volume.

The near-infrared spectra of T-dwarf stars were illustrated in Fig. 1. At first inspection, the potential for misidentification as line emission appears small.  The flux of a late-T or early-Y dwarf is peaky, depleted by deep molecular absorption lines. While these peaks become narrower and sharper with increasing spectral class \citep[a trend which continues through to the early Y-dwarfs,][]{2012ApJ...756..172M}, the remaining peaks span $\sim$1000\AA\ and the strongest (and thus most likely to peak up above the noise for faint sources) lies at $\lambda\sim1.25$\,$\mu$m. Since this corresponds to a Lyman $\alpha$ redshift of 9.3, it is unlikely to be taken at face value and may well conflict with the observed photometry.

\begin{figure*}
\includegraphics[width=\textwidth]{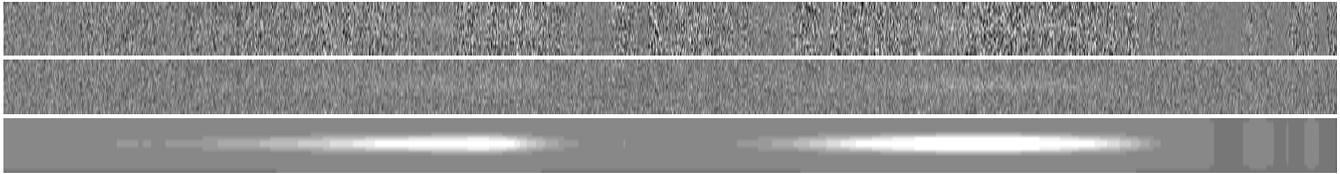}
\caption{The two dimensional spectrum of the T8 near-infrared standard star 2MASSI J0415195-093506, between 0.9 and 1.4 microns, normalized to the expected flux for a J$_{AB}=27.5$ source. We distribute the flux over a 0.8$''$ seeing disc in the spatial direction and assume 0.24$''$ pixels (as in, for example, XShooter). In the bottom panel, we show the predicted flux of the star before observational uncertainties. The second panel perturbs this with a random noise distribution consistent with that measured between sky emission lines at this wavelength by \citet{2012MNRAS.427.3055C} in a 4.9\,hr on-source observation using XShooter. In the top panel, we further add the Poisson-noise residuals of night sky emission OH lines, to demonstrate regions in which line identification is challenging or impossible. Wavelength increases towards the right, and the source is centred vertically in the virtual slit. \label{fig:T8_2d}}
\end{figure*}

\begin{figure*}
\includegraphics[width=\textwidth]{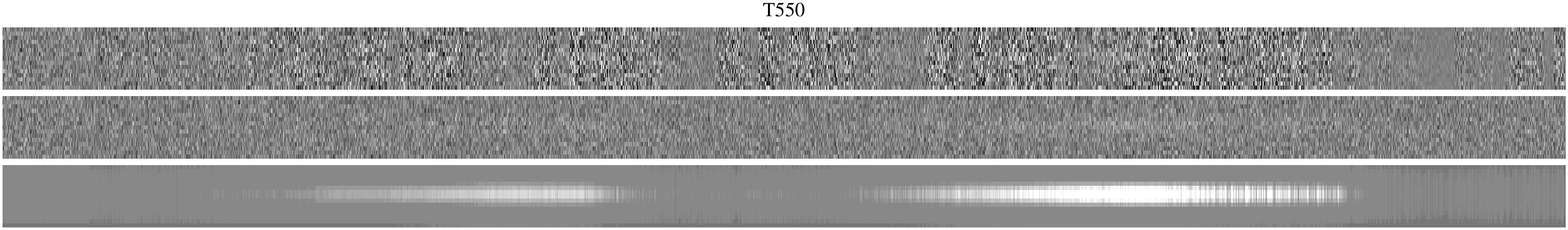}
\caption{As in Fig. \ref{fig:T8_2d} but for a 550\,K, $f_{sed}=4$, log($g$)=5.0 sulphide cloud model found to match the observed photometry of Y-dwarfs \citep[][]{2012ApJ...756..172M}.\label{fig:500K_2d}}
\end{figure*}

However, in figures \ref{fig:T8_2d} and \ref{fig:500K_2d} we illustrate the effects of a plausible observing strategy on two cool dwarf spectra - the class T8 near-infrared standard object, 2MASSI J0415195-093506, taken from the SpeX Prism Spectral Library\footnote{Available at \url{http://pono.ucsd.edu/~adam/browndwarfs/spexprism/index.html}} and a 550\,K, $f_{sed}=4$, log($g$)=5.0 sulphide cloud model found to match the observed photometry of Y-dwarfs \citep[][]{2012ApJ...756..172M}. We model our simulation on the strategy employed by \citet{2012MNRAS.427.3055C} on their source `ERS.YD2', using the XShooter instrument on the ESO/VLT. We normalize the source flux to J$_{AB}=27.5$, a typical magnitude for detected $z\sim7$ candidates in the CANDELS GOODS-N field \citep{elbert}. This flux is distributed over a 0.8$''$ seeing disc in the spatial direction and we assume a spatial scale of 0.24$''$ per pixel. In the bottom panel of each Fig., we show the predicted flux of the star before observational uncertainties. The second panel perturbs this with a random noise distribution consistent with that measured between sky emission lines at this wavelength by \citet{2012MNRAS.427.3055C} in a 4.9\,hr on-source observation. In the top panel, we further add the Poisson-noise residuals expected to be left after careful subtraction of night sky emission OH lines, derived from the measured night sky in the infrared at Cerro Pachon using Gemini-South\footnote{see \url{http://www.gemini.edu/sciops/telescopes-and-sites/observing-condition-constraints}, 2.3mm of water vapour}, to demonstrate regions of the spectrum in which line identification is challenging or impossible.

As these figures (which span the first two flux peaks in the near-infrared) demonstrate, for a cool dwarf star at zero radiative velocity relative to the Sun the second flux peak, at $\sim1.25$\,$\mu$m, is coincident with a strong complex of emission lines (and thus residuals in reduced data) arising from night sky emission.  This throws into question whether flux detected from this peak in moderate resolution (R of a few thousand) spectroscopy would be recognized or whether it would be dismissed as likely sky subtraction residuals.  If anything, the probability of the latter scenario is higher since the presence of underlying flux may well compromise the sky subtraction in this complex, leading to still larger noise residuals at these wavelengths. Certainly, it is clear that any line interpreted as Lyman $\alpha$ at $z\sim9.3$ must be investigated with extreme caution.

\begin{figure*}
\includegraphics[width=\textwidth]{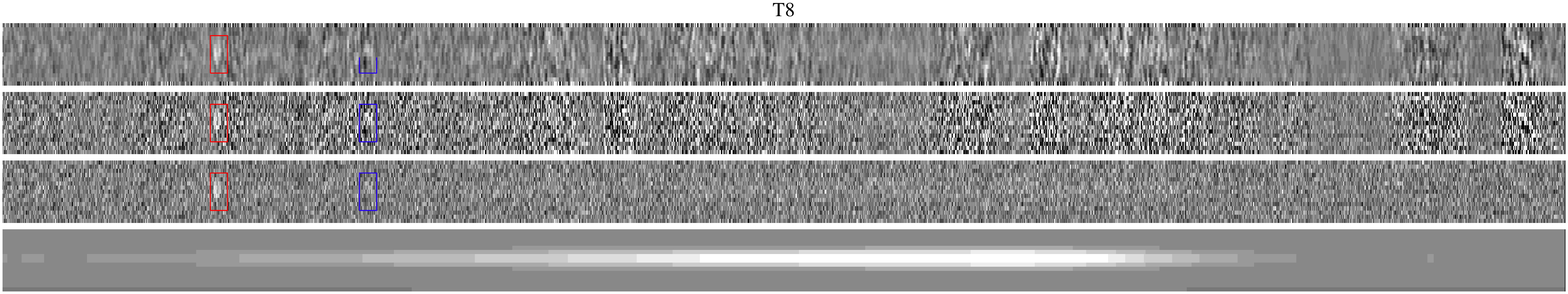}
\caption{As in Fig. \ref{fig:T8_2d}, but now focussing on $0.9<\lambda<1.2$\,$\mu$m, which corresponds to Lyman $\alpha$ redshifts $6.4<z<8.8$. At this resolution, it is possible to begin to see individual features appearing above the noise. The bottom panel shows the raw spectrum, followed by the addition of random noise, noise plus sky lines, and finally a panel with noise and sky, smoothed on a scale of 3\AA\ to simulate analysis methods. To illustrate the expected signal, we inject a source with the properties of a formal 5\,$\sigma$ emission line detection with a velocity width of 200\,km\,s$^{-1}$ towards the left-hand side of the upper panels, and mark the location with a red box. The blue box indicates the presence of a spurious `detection' at slightly longer wavelengths. \label{fig:T8_2dzoom}}
\end{figure*}

\begin{figure*}
\includegraphics[width=\textwidth]{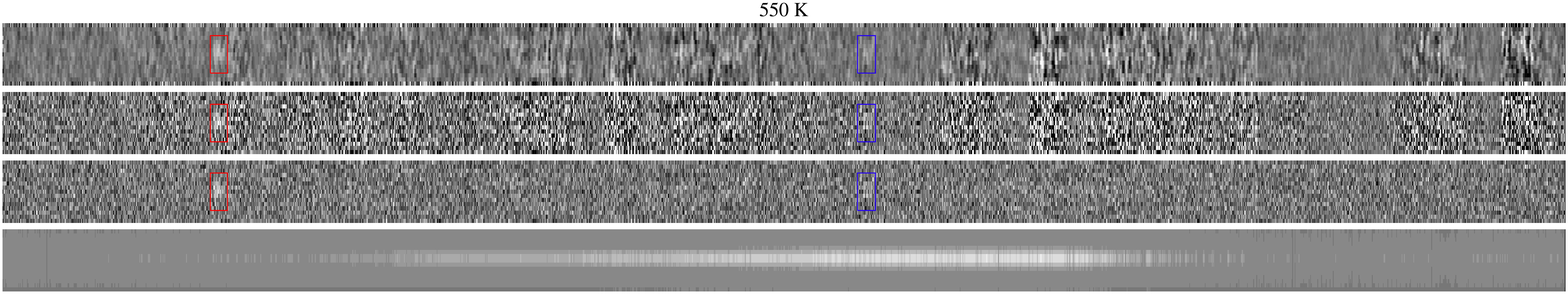}
\caption{As in Fig. \ref{fig:T8_2dzoom}, but for the 550\,K Y-dwarf model discussed in the text.  \label{fig:500K_2dzoom}}
\end{figure*}

Given that the presence of the stronger peak may well be overlooked or ambiguous, we turn our attention instead to the first spectral peak in the infrared spectra of late-T/early-Y dwarfs -- that occurring at $\lambda\sim1.08$\,$\mu$m -- in figures \ref{fig:T8_2dzoom} and \ref{fig:500K_2dzoom}. At this resolution, it is possible to begin to see individual features appearing above the noise.  To illustrate the expected signal, we inject a source with the properties of a formal 5\,$\sigma$, asymmetric emission line detection with a velocity width of 200\,km\,s$^{-1}$ as might be expected for higih-redshift Lyman $\alpha$ emission. We mark the location of this false source towards the left-hand side of the upper panels, where the uppermost panel is slightly smoothed, as might be expected to improve the recovery of genuine signals during analysis. 

\begin{figure}
\includegraphics[width=\columnwidth]{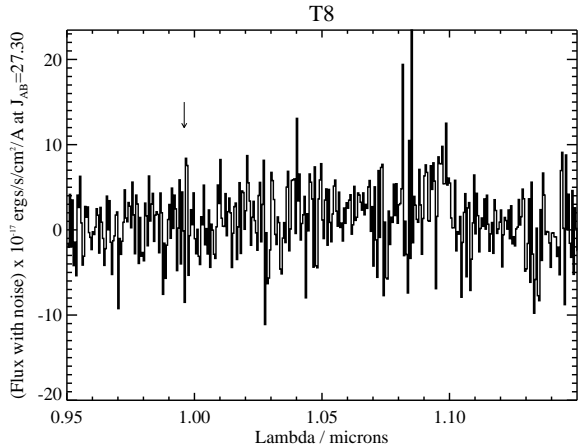}
\caption{An extracted spectrum from the realization in Fig. \ref{fig:T8_2dzoom}, rebinned from a native resolution of 1\AA\ (as for XShooter) to 4\AA\ bins for clarity. The feature identified with an arrow exceeds the noise standard deviation by a factor of a few across a FWHM of 8\AA\ - precisely as expected for a higih-redshift Lyman $\alpha$ line at a signal-to-noise of $\sim7$. It is, nonetheless, spurious.\label{fig:T8_z}}
\end{figure}

\begin{figure}
\includegraphics[width=\columnwidth]{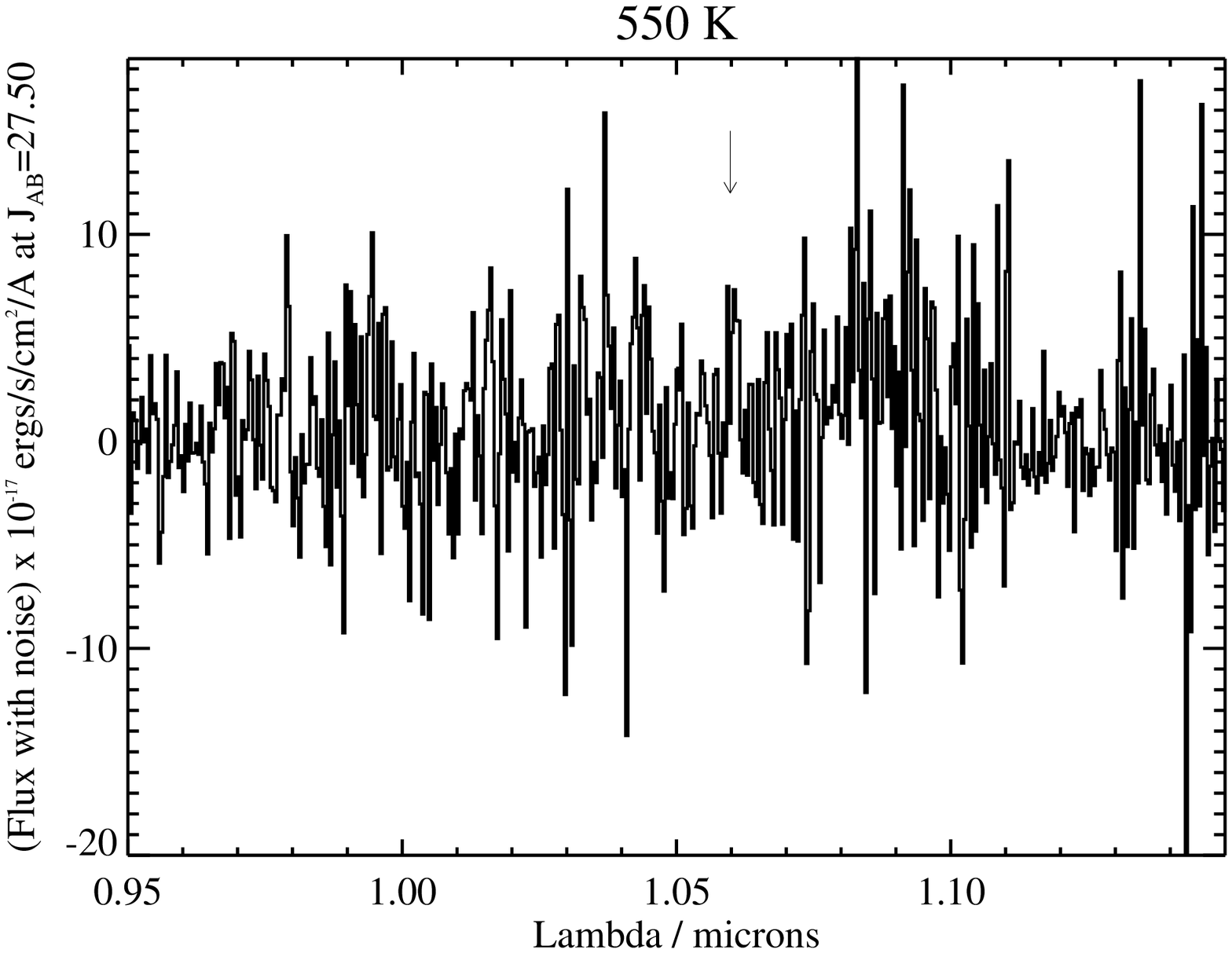}
\caption{As in Fig. \ref{fig:T8_z}, but for the 550\,K Y-dwarf model discussed in the text. \label{fig:500K_z}}
\end{figure}

The blue box in each Fig. indicates the presence of a spurious `detection' at slightly longer wavelengths. Interestingly, these are not coincident with the very peak of stellar emission, but rather appear to be a result of the broad wings of the stellar flux peak boosting random noise fluctuations above the detection threshold. This may, in fact, be the dominant mode by which late-T and early-Y dwarfs contaminate spectroscopic samples. The broad peaks in their spectra boost the background flux over several thousand angstroms, but are themselves likely to lie below the continuum detection limit. As a result, their presence, and the subsequent effect on the probability of spurious detections can be extremely hard to interpret, particularly if the smooth flux distribution affects background subtraction.  As can be seen from figures \ref{fig:T8_z} and \ref{fig:500K_z}, this effect can be equally difficult to detect in a 1D extracted spectrum. Features spanning $\sim$10\AA\ and lying well above the noise level can be identified, even between night sky emission lines.

Calculating the probability of a situation such as this occurring depends on many variables, most of them unknown. Late-T and early-Y dwarfs will peak at different frequencies as the relative strengths of molecular absorption bands combine to move peaks in their residual continuum emission. These peak wavelengths may be further perturbed by any significant radial velocity relative to the Sun, making them impossible to predict accurately, even given precise and unambiguous stellar models (which do not yet exist).

Furthermore, the depth and resolution of the spectrograph in question will play an important role. In our simulations with the Y-dwarf model, 20.4\% of noise realizations produced a 5\,$\sigma$ spurious `detection' not coincident with a strong sky line, and $>$99\% produced a 3\,$\sigma$ flux excess between 0.95 and 1.15\,$\mu$m picked up in the simulated extraction window (again, modelled on the observations of Caruana et al 2012, 2013). Deeper or shallower observations, brighter or fainter target sources, and different model parameters will all effect these fractions.  Very low resolution spectroscopy (as obtained, for example, by space telescope-mounted grisms), is likely to be less prone to uncertainty, since the continuum flux of the stars will be integrated to a detectable level, and the vagaries of sky emission variability will not affect the data. Sufficiently deep ground based spectroscopy \citep[a factor of $\sim3$ deeper than used by][]{2012MNRAS.427.3055C} would likely reveal broad underlying emission, or perhaps improve the detectability of the longer-wavelength flux peaks against the sky residual background. However, this effectively requires an increase by a factor of ten in the already long and thus expensive integrations with the world's largest near-infrared telescopes. For brighter targets, of course, identification of Y-dwarfs from spectroscopy may be easier to achieve. Unfortunately, bright targets are rare and require large areas to be surveyed for their identification. Given the observing expense involved, the $z>7$ candidate samples are more likely to grow through improved depth than increased area. The upcoming {\em Hubble} Frontier Fields, for example, will probe significantly deeper than the CANDELS survey, and the majority of CANDELS targets are already extremely difficult to observe spectroscopically.

While many of these factors are impossible to quantify in any realistic way, the simulation presented here demonstrates the difficulty confronting observers. Detection of a single emission line, particularly at a wavelength close to the peaks of the cool dwarf spectral emission, must be approached with caution. As a first check, heavy smoothing of the 2D spectrum may indicate a broad emission component at the target's spatial location on the slit. In addition, long slits are clearly preferable to short slitlets since the underlying continuum peaks are likely to render sky subtraction difficult across the near-infrared OH complexes, unless their spatial location can be isolated and omitted from any sky-line fitting procedure.

These recommendations cannot, of course, rule out misinterpretation of Y-dwarf flux as Lyman $\alpha$ emission and, should a sizeable population of these sources be found outside the local volume, this source of potential contamination should not be overlooked. We also note that much of this analysis is also relevant for the search for AGN in a similar redshift range. 

\section{Conclusions}\label{sec:conc}

It is now possible, using deep near-IR imaging, to identify and study both star-forming galaxies and active galactic nuclei at very-high ($z>6$) redshift. Unfortunately, cool and low-mass stars can exhibit similar broad-band colours potentially contaminating high-redshift selections. Our main conclusions can be summarized as follows:\\

\noindent
{\em (i)} It is possible to photometrically distinguish cool, low-mass stars from high-redshift galaxies at $z<7$ using {\em Hubble}/WFC3 or VISTA/VIRCam. However, at $z\sim 7.5$, where high-redshift galaxies are selected predominantly by their $Y-J$ and $J-H$ colours this becomes extremely challenging for all but the brightest sources.\\

\noindent
{\em (ii)} Using broad-band {\em JWST}/NIRCam observations the situation reverses somewhat. It becomes possible at $z>8$ (using $F115W$, $F150W$, and $F200W$ observations) but challenging at $z\sim 7$.\\

\noindent
{\em (iii)} The problematic separation of star-forming galaxies at $z\sim 7$ and low-mass stars can be overcome using a combination of {\em JWST} wide and medium-band ($F140M$, $F162M$) photometry by taking advantage of the deep H$_2$O and CH$_4$ absorption bands in the spectra of low-mass stars.\\

\noindent
{\em (iv)} Single emission line detections, often interpreted to be Lyman-$\alpha$ in low signal-to-noise spectroscopic observations may also be suspect, due to the unusual characteristics of the cool dwarf star population.  

\subsection*{Acknowledgements}

This research has benefitted from the SpeX Prism Spectral Libraries, maintained by Adam Burgasser at \url{http://pono.ucsd.edu/~adam/browndwarfs/spexprism}.

\end{document}